\documentclass[12pt]{article}
\def\bea{\begin{eqnarray}}
\def\eea{\end{eqnarray}}
\def\nn{\nonumber}
\def\lb{\label}
\def\half{\frac{1}{2}}
\def\lrb{\left(}
\def\lcb{\left\{}

\def\rrb{\right)}
\def\rcb{\right\}}

\def\C{{\cal C}}

\begin{document}
\begin{center}
{\Large On the $q$-analogues of the Zassenhaus formula \\
for disentangling exponential operators} \\

\bigskip

{\large R. Sridhar\footnote{{\em E-mail}: sridhar@imsc.res.in}
and R. Jagannathan\footnote{{\em E-mail}: jagan@imsc.res.in}} \\

\medskip

{\it The Institute of Mathematical Sciences \\
C.I.T. Campus, Tharamani, Chennai, TN 600113, India} \\

\bigskip

{\sl To our dear friend Srinivasa Rao with admiration, affection, and
best wishes \\
on his sixtieth birthday}

\end{center}

\bigskip

\noindent{\bf Abstract}

\smallskip

Katriel, Rasetti and Solomon introduced a $q$-analogue of the Zassenhaus
formula written as
$e_q^{(A+B)}$ $=$ $e_q^Ae_q^Be_q^{c_2}e_q^{c_3}e_q^{c_4}e_q^{c_5}\cdots$,
where $A$ and $B$ are two generally noncommuting operators and $e_q^z$
is the Jackson $q$-exponential, and derived the expressions for $c_2$,
$c_3$ and $c_4$.  It is shown that one can also write $e_q^{(A+B)}$ $=$
$e_q^Ae_q^Be_{q^2}^{\C_2}e_{q^3}^{\C_3}e_{q^4}^{\C_4}e_{q^5}^{\C_5}\cdots$.
Explicit expressions for $\C_2$, $\C_3$ and $\C_4$ are given.

\bigskip

\vspace{1cm}

\noindent
{\bf 1. Introduction}

\medskip

As is well known, the Baker-Campbell-Hausdorff (BCH) formula
\bea
e^Ae^B & = & e^{A+B+\half[A,B]+\frac{1}{12}([A,[A,B]]+[[A,B],B])+\cdots}
\lb{bch}
\eea
helps express the product of two noncommuting exponential operators as a
single exponential operator in which the exponent is, in general, an infinite
series in terms of repeated commutators and several hundred terms of the
series have been calculated using the computer. The dual of the BCH formula
is the Zassenhaus formula
\bea
e^{A+B} & = & e^Ae^Be^{-\half[A,B]}
              e^{\frac{1}{6}[A,[A,B]]-\frac{1}{3}[[A,B],B]}\cdots
\lb{zass}
\eea
which helps disentangle an exponential operator into a product of, in general,
an infinite series of exponential operators involving repeated commutators
(see \cite{M} for details).  The BCH and the Zassenhaus formulas have
several applications (see, for example, (\cite{W}-\cite{H})).

With the advent of $q$-deformed algebraic structures in physics (see, for
example, \cite{B,C,J} and references therein) there has been a growing
interest in $q$-generalization of several results of classical analysis.
Katriel and Solomon \cite{K1} first obtained the $q$-analogue of the BCH
formula and later Katriel, Rasetti and Solomon \cite{K2} proposed a
$q$-analogue of the Zassenhaus formula.  Let us briefly recall their results.

The Jackson $q$-exponential is defined by
\bea
e_q^x & = & \sum_{n=0}^\infty \frac{x^n}{[n]!}
\eea
where
\bea
[n] & = & \frac{1-q^n}{1-q}
\lb{qdn}
\eea
is the Heine basic number and
\bea
[n]! & = & [n][n-1][n-2]\cdots[1] \qquad
[0]! = 1
\eea
such that
\bea
\lim_{q\rightarrow 1}\,[n] & = & n \qquad
\lim_{q\rightarrow 1}\,e_q^x = e^x.
\eea
In the following $[n]$ will refer to the $q$-deformed $n$ defined by
(\ref{qdn}) corresponding to the base $q$.  If the base is different then
it will be indicated explicitly; for example, $[n]_{q^k}$ will mean
$(1-q^{kn})/(1-q^k)$.  The $q$-exponential function $e_q^{\alpha x}$ is
the eigenfunction of the Jackson $q$-differential operator
\bea
D_qf(x) & = & \frac{f(x)-f(qx)}{(1-q)x}
\eea
such that
\bea
D_qe_q^{\alpha x} & = & \alpha e_q^{\alpha x}.
\eea
The $q$-commutator is defined by
\bea
[X,Y]_q & = & XY-qYX
\eea
and it obeys the $q$-antisymmetry property
\bea
[Y,X]_q & = & -q[X,Y]_{q^{-1}}.
\eea

The $q$-BCH formula found in \cite{K1} reads
\bea
e_q^Ae_q^B & = & e_q^{A+B+\frac{q}{[2]}[A,B]_{q^{-1}}
                 +\frac{q^2}{[2]3!}\lrb[A, [A,B]_q]_{q^{-1}}
                 +[[A,B]_q,B]_{q^{-1}}\rrb+\cdots}.
\lb{qbch}
\eea
Katriel and Solomon \cite{K1} have given the explicit expressions for the
terms involving up to 4-tuple $q$-commutator.  In the limit $q$
$\longrightarrow$ $1$ the above $q$-BCH formula (\ref{qbch}) is seen to
agree with the classical BCH formula (\ref{bch}).

When $A$ and $B$ satisfy the relation $AB$ $=$ $q^{-1}BA$ it is found that
$[A,B]_{q^{-1}}$, $[A,[A,B]_q]_{q^{-1}}$, $[[A,B]_q,B]_{q^{-1}}$, and all
the higher $q$-commutators vanish leading to the result of Sch\"{u}tzenberger
\cite{S} and Cigler \cite{Ci} (see also, \cite{F})\,:
\bea
e_q^Ae_q^B & = & e_q^{A+B} \qquad \mbox{if}\ AB = q^{-1}BA.
\lb{sc}
\eea

The $q$-analogue of the Zassenhaus formula proposed in \cite{K2} reads
\bea
e_q^{(A+B)} = e_q^Ae_q^Be_q^{c_2}e_q^{c_3}e_q^{c_4}e_q^{c_5}\cdots
\lb{krs}
\eea
where
\bea
c_2 & = & [B,A]_q/[2], \nn \\
c_3 & = & ([[B,A]_q,A]_{q^2}/[3]!)+([[B,A]_q,B]_q/[3]), \nn \\
c_4 & = & ([[[B,A]_q,A]_{q^2},A]_{q^3}/[4]!)
           +([[[B,A]_q,B]_{q^2},B]_{q^3}/[2][4]) \nn \\
    &   &  +([[[B,A]_q,A]_{q^2},B]_q/[2][4])
           +(q[[B,A]_q,[B,A]_q]_q/[2]^2[4]), \nn \\
    &   & \ldots\,.
\eea

In this article we shall see that it is possible to have a $q$-Zassenhaus
formula written also as
\bea
e_q^{(A+B)} = e_q^Ae_q^Be_{q^2}^{\C_2}
              e_{q^3}^{\C_3}e_{q^4}^{\C_4}e_{q^5}^{\C_5}\cdots
\eea
and give the explicit expressions for the first few terms.  \\

\noindent
{\bf 2. The Zassenhaus formula}

\medskip

\noindent
A standard procedure to obtain the Zassenhaus formula is as follows.  Set
\bea
e^{x(A+B)} & = & e^{xA}e^{xB}e^{x^2C_2}e^{x^3C_3}\cdots\,.
\lb{exa+b}
\eea
Differentiating both sides of (\ref{exa+b}) with respect to $x$ and
multiplying it from the right by
\bea
e^{-x(A+B)} & = & \cdots e^{-x^3C_3}e^{-x^2C_2}e^{-xB}e^{-xA}
\eea
one obtains
\bea
A+B & = & A+e^{xA}Be^{-xA}+e^{xA}e^{xB}(2xC_2)e^{-xB}e^{-xA} \nn \\
    &   & \quad +e^{xA}e^{xB}e^{x^2C_2}(3x^2C_3)e^{-x^2C_2}e^{-xB}e^{-xA}
         +\cdots.
\lb{a+b}
\eea
The expressions $e^{xA}Be^{-xA}$, $e^{x}e^{xB}(2xC_2)e^{-xB}e^{-xA}$, etc.,
are expanded again using the formula
\bea
e^{xA}Be^{-xA} & = & \sum_{n=0}^\infty \frac{x^n}{n!}\{A^n,B\}
\lb{eabe-a}
\eea
where the multiple commutator bracket $\{A^n,B\}$ is defined by
\bea
\{A^{n+1},B\} & = & [A,\{A^n,B\}] \quad \{A^0,B\} = B.
\eea
Then, equation (\ref{a+b}) becomes
\bea
0 & = & \sum_{n=1}^\infty\frac{x^n}{n!}\{A^n,B\} \nn \\
  &   & +2x\sum_{m=0}^\infty\sum_{n=0}^\infty\frac{x^{m+n}}{m!n!}
        \{A^m,B^n,C_2\} \nn \\
  &   & +3x^2\sum_{l=0}^\infty\sum_{m=0}^\infty\sum_{n=0}^\infty
        \frac{x^{l+m+2n}}{l!m!n!}\{A^l,B^m,C_2^n,C_3\}+\cdots
\lb{0=}
\eea
with
\bea
\{A^0,B^n,C_2\} & = & \{B^n,C_2\} \nn \\
\{A^{m+1},B^n,C_2\} & = & [A,\{A^m,B^n,C_2\}] \qquad \cdots .
\eea
Equating the coefficients of $x^n$ to zero in (\ref{0=}) one obtains
\bea
C_2 & = & -\half[A,B] \nn \\
C_3 & = & \frac{1}{6}[A,[A,B]]-\frac{1}{3}[[A,B],B] \qquad \cdots\,.
\eea

There are also alternative methods of derivation of the above result
(see, for example, \cite{W,Wi} for more details).  For our purpose of
deriving a $q$-analogue of the Zassenhaus formula we shall follow
Karplus and Schwinger \cite{Ka} (see Appendix I where $\exp(A+B)$ is
expanded in powers of $B$ up to the second term using a method we adopt
here).  \\

\noindent
{\bf 3. A $q$-analogue of the Zassenhaus formula}

\medskip

\noindent
Let
\bea
F(x) & = & e_q^{x(A+B)} \qquad F(0) = I
\eea
and write
\bea
e_q^{x(A+B)} & = & e_q^{xA}G(x) \qquad G(0) = I.
\lb{f=eg}
\eea
On $q$-differentiating with respect to $x$
\bea
D_qF(x) & = & (A+B)e_q^{x(A+B)}.
\eea
Note that $q$-differentiation obeys the $q$-Leibniz rule
\bea
D_q(fg) & = & (D_qf)g+f(qx)(D_qg) = (D_qf)g(qx)+f(D_qg).
\eea
Now, $q$-differentiating the right hand side of (\ref{f=eg}) using the
$q$-Leibniz rule we have
\bea
(A+B)e_q^{x(A+B)} & = & Ae_q^{xA}G(x)+e_q^{qxA}D_qG(x) \nn \\
                  & = & Ae_q^{x(A+B)}+e_q^{qxA}D_qG(x).
\eea
Thus,
\bea
Be_q^{x(A+B)} & = & Be_q^{xA}G(x) = e_q^{qxA}D_qG(x).
\eea
Or
\bea
D_qG(x) & = & (e_q^{qxA})^{-1}Be_q^{xA}G(x).
\lb{dqg}
\eea

Recall that the inverse of $q$-differentiation is $q$-integration (see,
for example, \cite{E,G}), defined
by
\bea
\int_0^x d_q\xi f(\xi) & = & (q-1)x\sum_{n=1}^\infty q^{-n}f(q^{-n}x)
\eea
such that
\bea
D_q\lrb\int_0^x d_q\xi f(\xi)\rrb & = & f(x).
\eea
Then, the formal solution of the $q$-differential equation (\ref{dqg})
for $G(x)$ with the initial condition $G(0)$ $=$ $I$, is given by
\bea
G(x) & = & I+\int_0^x d_q\xi(e_q^{q\xi A})^{-1}Be_q^{\xi A}G(\xi) \nn \\
     & = & I+\int_0^x d_q\xi(e_q^{q\xi A})^{-1}Be_q^{\xi A} \nn \\
     &   & \quad +\int_0^x d_q\xi_1(e_q^{q\xi_1A})^{-1}Be_q^{\xi_1A}
                  \int_0^{\xi_1} d_q\xi_2(e_q^{q\xi_2A})^{-1}Be_q^{\xi_2A}
                  \nn \\
     &   & \quad +\cdots\,.
\lb{formalg}
\eea
Using the well known result
\bea
e_q^xe_{q^{-1}}^{-x} & = & 1
\eea
we can rewrite (\ref{formalg}) as
\bea
G(x) & = & I+\int_0^x d_q\xi e_{q^{-1}}^{-q\xi A}Be_q^{\xi A} \nn \\
     &   & \quad +\int_0^x d_q\xi_1 e_{q^{-1}}^{-q\xi_1 A}Be_q^{\xi_1A}
                  \int_0^{\xi_1} d_q\xi_2 e_{q^{-1}}^{-q\xi_2 A}Be_q^{\xi_2A}
                  \nn \\
     &   & \quad +\cdots\,.
\lb{g}
\eea

We now require a $q$-analogue of the classical formula (\ref{eabe-a}).
By straightforward expansion and regrouping of terms in powers of $x$ we
have
\bea
(e_q^{qxA})^{-1}Be_q^{xA} & = & e_{q^{-1}}^{-qxA}Be_q^{xA} \nn \\
 & = & B+x[B,A]_q+\frac{x^2}{[2]!}[[B,A]_q,A]_{q^2} \nn \\
 &   & \quad +\frac{x^3}{[3]!}[[[B,A]_q,A]_{q^2},A]_{q^3}+\cdots\,.
\eea
Let us write this equation as
\bea
e_{q^{-1}}^{-qxA}Be_q^{xA} & = & B+\sum_{n=1}^\infty\frac{x^n}{[n]!}X_n
\lb{eqa-ba}
\eea
with the definition
\bea
X_n & = & [\cdots[[[B,A]_q,A]_{q^2},A]_{q^3}\cdots],A]_{q^n}
          \quad n = 1,2,3,\cdots\,.
\lb{x}
\eea

Now, using (\ref{eqa-ba}) and (\ref{x}) in (\ref{g}), we get
\bea
G(x) & = & I+\int_0^x d_q\xi\lcb B
            +\sum_{n=1}^\infty\frac{\xi^n}{[n]!}X_n\rcb \nn \\
     &   &  +\int_0^x d_q\xi_1\lcb B
            +\sum_{n=1}^\infty\frac{\xi_1^n}{[n]!}X_n\rcb
            \int_0^{\xi_1}d_q\xi_2\lcb B
            +\sum_{n=1}^\infty\frac{\xi_2^n}{[n]!}X_n\rcb \nn \\
     &   &  +\cdots\,.
\eea
Collecting the first few terms of the resulting series in powers of $x$ we
have
\bea
G(x) & = & \lcb I+xB+\frac{x^2B^2}{[2]!}+\frac{x^3B^3}{[3]!}+\cdots\rcb \nn \\
     &   & \quad +\lcb\frac{x^2X_1}{[2]!}+\frac{x^4X_1^2}{[1][2]![4]}
                     +\frac{x^6X_1^3}{[1]^2[2]![4][6]}+\cdots\rcb \nn \\
     &   & \quad +\lcb x^3\lrb\frac{X_2}{[3]!}+\frac{BX_1}{[3]!}
                 +\frac{X_1B}{[1][3]}\rrb+x^6\lrb \cdots \rrb+\cdots\rcb \nn \\
     &   & \quad +\cdots\,.
\lb{gx}
\eea
Realizing that the terms in the first bracket sum to $e_q^{xB}$ let us rewrite
(\ref{gx}) as
\bea
G(x) & = & e_q^{xB}\lrb I
            +e_{q^{-1}}^{-xB}\lcb\frac{x^2X_1}{[2]!}
            +\frac{x^4X_1^2}{[1][2]![4]}+\frac{x^6X_1^3}{[1]^2[2]![4][6]}
            +\cdots\rcb \right. \nn \\
     &   & \quad \left. +e_{q^{-1}}^{-xB}\lcb x^3\lrb\frac{X_2}{[3]!}
              +\frac{BX_1}{[3]!}+\frac{X_1B}{[1][3]}\rrb+x^6\lrb\cdots\rrb
              +\cdots\rcb +\cdots\rrb. \nn \\
     &   &
\lb{gx2}
\eea
Substituting the series expansion $e_{q^{-1}}^{-xB}$, using the relation
\bea
[n]_{q^{-1}} & = & q^{1-n}[n]_q
\eea
and after some straightforward algebra one can rewrite (\ref{gx2}) further as
\bea
G(x) & = & e_q^{xB}e_{q^2}^{x^2X_1/[2]}\lrb I+e_{q^2}^{-x^2X_1/[2]}
           \lcb -xB+\frac{qx^2B^2}{[2]}-\frac{q^2x^3B^3}{[3]}
           +\cdots\rcb\cdots \right. \nn \\
     &   & \quad \left. +e_{q^2}^{-x^2X_1/[2]}\lcb x^3\lrb\frac{X_2}{[3]!}
           +\frac{BX_1}{[3]!}+\frac{X_1B}{[3]}\rrb +\cdots\rcb+\cdots\rrb.
\eea
Now, substituting the series expression for $e_{q^2}^{-x^2X_1/[2]}$ and
simplifying, one recognizes that one can pull out a factor
$e_{q^3}^{x^3\{([X_1,B]_q/[3])+(X_2/[3]!)\}}$ from the above expression and
write
\bea
G(x) & = & e_q^{xB}e_{q^2}^{x^2[B,A]_q/[2]}
           e_{q^3}^{x^3\{([[B,A]_q,B]_q/[3])
              +([[B,A]_q,A]_{q^2}/[3]!)\}}\cdots \nn \\
     & = & e_q^{xB}e_{q^2}^{-qx^2[A,B]_{q^{-1}}/[2]}
           e_{q^3}^{x^3\{(q^3[A,[A,B]_{q^{-1}}]_{q^{-2}}/[3]!)
              -(q[[A,B]_{q^{-1}},B]_q/[3])\}}\cdots\,. \nn \\
     &   &
\lb{eee}
\eea
This shows that the general expression for $G(x)$ can be assumed to be
\bea
G(x) & = &
e_q^{xB}e_{q^2}^{x^2\C_2}e_{q^3}^{x^3\C_3}e_{q^4}^{x^4\C_4}\cdots\,.
         \nn \\
     &   &
\lb{ccc}
\eea
The crucial point here is to note that the exponential factors in $G(x)$
have bases $q$, $q^2$, $q^3$, $q^4$, $\cdots$, respectively, unlike in
the formula (\ref{krs}) introduced by Katriel, Rasetti and Solomon \cite{K2}.
Thus, having recognized a new general form of $G(x)$ we can use the
comparison method to determine $\C_2$, $\C_3$, $\C_4$, $\cdots$, in
(\ref{ccc}).  To this end, we write (\ref{f=eg}) as
\bea
\sum_{n=0}^\infty\frac{x^n}{[n]!}(A+B)^n & = &
\sum_{j,k,l,m,\cdots=0}^\infty
\frac{x^{j+k+2l+3m+\cdots}}{[j]_q![k]_q![l]_{q^2}![m]_{q^3}!}
A^jB^k\C_2^l\C_3^m\cdots
\eea
and compare the coefficients of equal powers of $x$.  This leads to coupled
equations for $\C_2$, $\C_3$, $\C_4$, $\cdots$, which can be solved in
terms of $A$ and $B$ after straightforward algebra.  Thus, solving the
first few equations we confirm the forms of $\C_2$ and $\C_3$ already
obtained above (compare (\ref{eee}) and (\ref{ccc})) and find that
\bea
\C_4 & = & \frac{1}{[4]!}\lrb -q^6[A,[A,[A,B]_{q^{-1}}]_{q^{-2}}]_{q^{-3}}
          \right. \nn \\
    &   & \quad \left. +q^3[3][[A,[A,B]_{q^{-1}}]_{q^{-2}},B]_q
                       -q[3][[[A,B]_{q^{-1}},B]_q,B]_{q^2}\rrb.
\lb{c4}
\eea
Now, using (\ref{f=eg}), (\ref{eee}), (\ref{ccc}) and (\ref{c4}), we get
the $q$-Zassenhaus formula as
\bea
 &   &  e_q^{(A+B)} = e_q^Ae_q^Be_{q^2}^{-q[A,B]_{q^{-1}}/[2]}
                  e_{q^3}^{\{(q^3[A,[A,B]_{q^{-1}}]_{q^{-2}}/[3]!)
                  -(q[[A,B]_{q^{-1}},B]_q/[3])\}} \nn \\
 &   & \quad\times
                  e_{q^4}^{\frac{1}{[4]!}\lrb
                  -q^6[A,[A,[A,B]_{q^{-1}}]_{q^{-2}}]_{q^{-3}}
                  +q^3[3][[A,[A,B]_{q^{-1}}]_{q^{-2}},B]_q
                  -q[3][[[A,B]_{q^{-1}},B]_q,B]_{q^2}\rrb} \nn \\
 &   &
\lb{qzass}
\eea
up to the first five terms.  In the limit $q$ $\longrightarrow$ $1$ it is
found that $\C_2$, $\C_3$ and $\C_4$ become the expressions for the
classical Zassenhaus formula (see, for example, (7)-(9) of \cite{Wi}).
When $AB$ $=$ $q^{-1}BA$ it is found that $\C_2$ $=$ $\C_3$ $=$ $\C_4$
$=$ $\cdots$ $=$ $0$ leading to the Sch\"{u}tzenberger-Cigler result
(\ref{sc}).

In the literature on $q$-series there are two other definitions of the
$q$-exponential:
\bea
e_q(x) & = & \sum_{n=0}^\infty \frac{x^n}{(1-q)(1-q^2)\cdots(1-q^n)}
\lb{smalleq} \\
E_q(x) & = & \sum_{n=0}^\infty \frac{q^{n(n-1)/2}x^n}
             {(1-q)(1-q^2)\cdots(1-q^n)}.
\lb{bigeq}
\eea
such that
\bea
\lim_{q\rightarrow 1}\,e_q((1-q)x) & = & e^x \\
\lim_{q\rightarrow 1}\,E_q((1-q)x) & = & e^x \\
e_q(x)E_q(-x) & = & 1.
\eea
Let us note that the corresponding Zassenhaus formulae can be found in
straightforward ways by using the relations:
\bea
e_q(x) & = & e_q^{x/(1-q)}
\lb{eqeq} \\
E_q(x) & = & e_{q^{-1}}^{x/(1-q)}.
\lb{Eqeq}
\eea
Thus, we find, using (\ref{qzass}) and (\ref{eqeq}),
\bea
e_q(A+B) & = & e_q^{(A+B)/(1-q)} \nn \\
         & = & e_q(A)e_q(B)e_{q^2}(-q[A,B]_{q^{-1}}/(1-q))\cdots
\eea
and
\bea
e_q(A+B) & = & e_q(A)e_q(B) \quad \mbox{if}\ \ AB = q^{-1}BA.
\eea
Similarly,
\bea
E_q(A+B) & = & e_{q^{-1}}^{(A+B)/(1-q)} \nn \\
         & = & E_q(A)E_q(B)E_{q^2}(-[A,B]_q/(1-q))\cdots
\eea
and
\bea
E_q(A+B) & = & E_q(A)E_q(B) \quad \mbox{if}\ \ AB = qBA.
\eea

\noindent
{\bf 4. Conclusion}

\medskip

\noindent
To summarize, it is found that the classical Zassenhaus formula
\bea
e^{A+B} & = & e^Ae^Be^{-\half[A,B]}
              e^{\frac{1}{6}[A,[A,B]]-\frac{1}{3}[[A,B],B]} \nn \\
        &   & \quad\times e^{\frac{1}{4!}
                          \lrb -[A,[A,[A,B]]]+3[[A,[A,B]],B]
                           -3[[[A,B],B],B]\rrb}\cdots
\eea
has a $q$-analogue given by
\bea
 &   &  e_q^{(A+B)} = e_q^Ae_q^Be_{q^2}^{-q[A,B]_{q^{-1}}/[2]}
                  e_{q^3}^{\{(q^3[A,[A,B]_{q^{-1}}]_{q^{-2}}/[3]!)
                  -(q[[A,B]_{q^{-1}},B]_q/[3])\}} \nn \\
 &   & \quad\times
                  e_{q^4}^{\frac{1}{[4]!}\lrb
                  -q^6[A,[A,[A,B]_{q^{-1}}]_{q^{-2}}]_{q^{-3}}
                  +q^3[3][[A,[A,B]_{q^{-1}}]_{q^{-2}},B]_q
                  -q[3][[[A,B]_{q^{-1}},B]_q,B]_{q^2}\rrb} \nn \\
 &   & \quad\times \cdots
\lb{q-z}
\eea
where the $q$-exponential is defined by
\bea
e_q^x & = & \sum_{n=0}^\infty \frac{x^n}{[n]!}.
\lb{qexp}
\eea
Thus, we have shown that while Katriel, Rasetti and Solomon \cite{K2}
have proposed a $q$-analogue of the Zassenhaus formula in the form
\bea
e_q^{(A+B)} = e_q^Ae_q^Be_q^{c_2}e_q^{c_3}e_q^{c_4}e_q^{c_5}\cdots\,,
\eea
it is possible to have a $q$-Zassenhaus formula written also as
\bea
e_q^{(A+B)} = e_q^Ae_q^Be_{q^2}^{\C_2}
              e_{q^3}^{\C_3}e_{q^4}^{\C_4}e_{q^5}^{\C_5}\cdots\,.
\lb{qdef}
\eea
We have also explicitly found the first few terms of the disentanglement
formula (\ref{qdef}).

It should be noted that once a $q$-Zassenhaus formula is given for the
$q$-exponential defined by (\ref{qexp}) for the other two common
definitions of the $q$-exponential (\ref{smalleq}, \ref{bigeq}) found in
the literature the corresponding $q$-Zassenhaus formulae can be found
using the relations (\ref{eqeq}, \ref{Eqeq}).

\end{document}